\documentclass[12pt,a4paper]{article}
\usepackage[onehalfspacing]{setspace}
\usepackage{amsmath,mathtools,amssymb,amsthm,amsfonts}
\usepackage[authoryear]{natbib}
\usepackage{authblk}

\usepackage{algorithm}
\usepackage{algpseudocode}
\usepackage{caption,subcaption,tabularx,booktabs}
\usepackage[]{threeparttable}
\usepackage{rotating}
\usepackage[hmargin=1in,vmargin=1in]{geometry}

\theoremstyle{plain}

\theoremstyle{definition}

\providecommand{\definitionname}{Definition}
\providecommand{\theoremname}{Theorem}

\newcommand\fnote[1]{\captionsetup{font=footnotesize}\caption*{#1}}

\usepackage{comment,color}
 \includecomment{comment}
 \specialcomment{comment}{\begingroup\color{red}}{\endgroup}  

\title{Hybrid unadjusted Langevin methods for high-dimensional latent variable models}
\date{\today}
\author{Rub\'en Loaiza-Maya\thanks{ Rub\'en Loaiza-Maya gratefully acknowledges
support by the Australian Research Council through grant DE230100029.}}
\author{Didier Nibbering\thanks{Correspondence to: Department of Econometrics \& Business Statistics, Monash University, Clayton VIC 3800, Australia, e-mail: \textsf{didier.nibbering@monash.edu}}}
\author{Dan Zhu}
\affil{\small Department of Econometrics and Business Statistics, Monash University}

\begin{document}
\maketitle
\begin{abstract} 
\noindent 
The exact estimation of latent variable models with big data is known to be challenging. The latents have to be integrated out numerically, and the dimension of the latent variables increases with the sample size. This paper develops a novel approximate Bayesian method based on the Langevin diffusion process. The method employs the Fisher identity to integrate out the latent variables, which makes it accurate and computationally feasible when applied to big data. In contrast to other approximate estimation methods, it does not require the choice of a parametric distribution for the unknowns, which often leads to inaccuracies. In an empirical discrete choice example with a million observations, the proposed method accurately estimates the posterior choice probabilities using only 2\% of the computation time of exact MCMC.

\end{abstract}

\thispagestyle{empty}
\clearpage
\setcounter{page}{1}

\section{Introduction}\label{sec:introduction}
Latent variable models are widely used in econometrics due to their ability to capture complex stylised facts of economic data. 
Recent examples include mixed effects models \citep{danaher2023optimal}, discrete choice models \citep{loaiza2022fast}, and state space models \citep{koopman2017intraday}.
At the same time, many modern econometric estimation problems involve a large data set, which often results in a large number of latent variables in these models. Estimation of these high-dimensional latent variable models is challenging. 

Large numbers of latent variables complicate both maximum likelihood and Bayesian estimation. Maximum likelihood estimation is often infeasible because it is not possible to analytically integrate the latent variables from the likelihood. Bayesian estimation deals with the latent variables by targeting the augmented posterior: the joint posterior distribution of the model parameters and the latent variables. To this end, exact Bayesian estimation methods, like Markov chain Monte Carlo (MCMC) methods, require generation of the latent variables. This typically induces high autocorrelation in the Markov chain, which makes exact Bayesian methods also computational infeasible for many big data problems.

A promising approximate Bayesian method from the machine learning \citep{vollmer2016exploration,hodgkinson2021implicit} and the statistics literature \citep{dalalyan2017theoretical,nemeth2021stochastic} is the unadjusted Langevin algorithm (ULA). 
To generate samples from an approximation to the posterior, it utilizes the Euler-Maruyama discretization to approximate a continuous-time Langevin process based on the gradient of the logarithm of the posterior density. This discretization requires a choice of step size, which can substantially improve the rate of convergence relative to exact MCMC methods, at the cost of an approximation bias that increases in the dimension of the posterior \citep{hodgkinson2021implicit}. 
When applied to latent variable models, ULA would target the augmented posterior. For big data problems the dimension of the augmented posterior is large, which makes the approximation error substantial and the method highly inaccurate \citep{durmus2017nonasymptotic}. Metropolis-adjusted Langevin algorithms (MALA) have been proposed to correct for this bias \citep{roberts1996exponential}, but these suffer from large computational costs and poor estimation results in high dimensions \citep{de2021efficient}.

Our main contribution is the development of the hybrid unadjusted Langevin algorithm (HULA), which is a computationally feasible and accurate estimation method for challenging high-dimensional latent variable models. HULA directly targets the marginal posterior of the model parameters rather than the posterior that is augmented with latent variables. Our method requires evaluation of an unbiased gradient estimate for the logarithm of the marginal posterior density, which we construct via the Fisher identity \citep{poyiadjis2011particle}. This unbiased estimate requires the gradient of the logarithm of the augmented posterior density to be available in closed form, and efficient generation from the exact conditional posterior distribution of the latent variables.

HULA has four main advantages. 
First, for an appropriate choice of step size, the rate of convergence can be substantially improved while the approximation error is small.
By targeting directly the marginal posterior density, the HULA approximation error does not increase with the sample size but only with the dimension of the parameter space. 
Second, the method samples all model parameters at once rather than in blocks, and only requires the evaluation of a gradient with respect to the model parameters instead of both the model parameters and the latents as in ULA. 
Third, HULA is applicable to a wide range of econometric models, for instance, state space models, limited dependent variable models, and random coefficient models.
Fourth, for certain models, the computational cost of HULA can be further reduced via the use of subsampling methods. Subsampling methods have also been shown to substantially improve the scalability of ULA in models without latent variables \citep{nemeth2021stochastic}.

We illustrate the potential of HULA in a big data discrete choice setting. We fit a multinomial probit model to more than a million observations on purchases from ten pasta brands. Exact MCMC methods for this model are known to suffer from high autocorrelation, due to the large number of latent utilities in the model. Compared to exact MCMC, HULA exhibits substantially faster convergence and higher effective sample sizes. At the same time, we observe that the estimates for the posterior choice probabilities of HULA and MCMC are indistinguishable, and that the loss in predictive accuracy is negligible. Since the multinomial probit model allows for subsampling, the computational costs can be further reduced by taking a random subsample of 20\% of the observations to estimate the gradient. The accuracy of the posterior choice probabilities and posterior predictive show almost no decline, using only 2\% of the computation time of exact MCMC.

The limitations of exact MCMC methods for the estimation of latent variable models has led to the development of other approximate Bayesian approaches. For instance, the econometrics literature has adopted variational Bayes (VB) methods in the estimation of state space models \citep{quiroz2018gaussian}, choice models \citep{loaiza2022fast}, and tobit models \citep{loaiza2022afast}. Instead of sampling from the exact posterior, VB calibrates an approximation to the posterior using fast optimization techniques. This approach requires the researcher to specify a class of approximating densities, which can be challenging for complex models with many latent variables. Moreover, VB is known to accurately estimate posterior means but underestimates posterior variance \citep{blei2017variational}. This may bias other objects of interests that are nonlinear transformations of the parameters, such as the choice probabilities from a multinomial probit model. This is in fact the case in our empirical application, where we demonstrate that HULA produces more accurate posterior choice probabilities than VB for some choice alternatives, while the estimation time of HULA with subsampling is of the same order.

The outline of this paper is as follows. Section~\ref{sec:method} introduces Bayesian estimation via Langevin dynamics. Section~\ref{sec:hula} introduces the hybrid unadjusted Langevin algorithm for the estimation of econometric models with a large number of latent variables. Section~\ref{sec:mnp} illustrates the method in a choice model with many latent variables. Finally, Section~\ref{sec:conclusion} concludes.  


\section{Bayesian estimation via Langevin dynamics}\label{sec:method}
\subsection{Setting}
Consider a data vector $y = (y_1^\top,\dots,y_n^\top)^\top$ with $n$ observations and a corresponding set of latent variables $z =  (z_1^\top,\dots,z_n^\top)^\top$, where $z$ is an $m_z$-dimensional vector. This paper considers parametric models that admit an augmented likelihood function of the form
\begin{align}
    p(y,z|\theta) = p(z|\theta)\prod_{i=1}^n p(y_i|z_i,\theta),
\end{align}
where $\theta$ denotes an $m_{\theta}$-dimensional parameter vector. This includes, for instance, state space models, limited dependent variable models, and random coefficients models. 

Bayesian estimation is concerned with the computation of the posterior distribution
\begin{equation} 
p(\theta|y) \propto  p(\theta)\int p(y,z|\theta)dz,\label{Eq:posterior}
\end{equation}
for a given set of prior beliefs on $\theta$ represented by the prior density $p(\theta)$. Generally, this posterior has an intractable analytical form as evaluation of $\int p(y,z|\theta)dz$ is often a complex integration problem. Instead, computation of \eqref{Eq:posterior} typically involves the use of MCMC methods that generate samples from the augmented posterior distribution 
\begin{equation}
p(\theta,z|y)\propto p(y,z|\theta)p(\theta),
\end{equation}
which automatically generates samples from the desired posterior $p(\theta|y)$. This approach requires generation from the conditional posterior distribution $p(z|y,\theta)$, which typically induces high autocorrelation in the Markov Chain. Hence, exact MCMC methods may require a large number of draws, which makes it computationally impractical for models with large sets of latent variables. 

\subsection{The unadjusted Langevin algorithm}
An unadjusted Langevin algorithm for \eqref{Eq:posterior} would be based on the $m$-dimensional Langevin stochastic differential equation
\begin{equation}
\begin{bmatrix} d\theta_{t} \\ dz_{t} \end{bmatrix}=\begin{bmatrix}\triangledown_{\theta}\log p(\theta_t,z_t|y) \\ \triangledown_{z}\log p(\theta_t,z_t|y)\end{bmatrix}dt+ \sqrt{2} dB_{t},\label{eq:langevin0}
\end{equation}
where $B_{t}$ is an $m$-dimensional Brownian motion process, with $m = m_z + m_{\theta}$. Given certain regularity conditions, the continuous-time dynamic process in \eqref{eq:langevin0} has been shown to have $p(\theta,z|y)$ as its invariant distribution \citep{roberts1996exponential,durmus2017nonasymptotic}. However, sampling from \eqref{eq:langevin0} is not feasible. 

The Langevin diffusion process can be approximated using the Euler-Maruyama discretization, which leads to the recursive formula
\begin{equation}
\begin{bmatrix} \theta_{k+1} \\ z_{k+1} \end{bmatrix}
=\begin{bmatrix} \theta_{k} \\ z_{k} \end{bmatrix}+\tau\begin{bmatrix}\triangledown_{\theta}\log p(\theta_k,z_k|y) \\ \triangledown_{z}\log p(\theta_k,z_k|y)\end{bmatrix}+\sqrt{2\tau}\epsilon_{k},\label{Eq:Discrete}
\end{equation}
where $\epsilon_{k}\sim N({0}_{m},I_{m})$ and $\tau$ is a scalar determining the step size of the discretization. The process of iteratively sampling from \eqref{Eq:Discrete} is referred to as ULA \citep{durmus2017nonasymptotic}. Since the discretization may induce a bias, this algorithm generates draws from an approximation to $p(\theta,z|y)$. The step size $\tau$ controls the trade-off between the rate of convergence of the sampling algorithm and its approximation error. 

Provided that the gradients in \eqref{Eq:Discrete} can be efficiently computed, ULA is faster than traditional MCMC algorithms. In each sampling iteration, all elements of $\theta$ and $z$ are sampled at once rather than in blocks. Moreover, at the cost of inducing more bias in the approximate posterior, ULA can also lead to faster convergence \citep{hodgkinson2021implicit}, and thus requiring less sample iterations. 

However, both the approximation error and the computational costs of ULA can be large. First, the total variation norm between the exact posterior and the ULA approximation increases in $m$ \citep{dalalyan2017theoretical,durmus2017nonasymptotic}. Although the approximation error can be corrected with a Metropolis-Hastings step \citep{roberts1996exponential}, this step is time consuming, deteriorates convergence properties, and may lead to poor estimation results \citep{de2021efficient}. 
Second, evaluation of the gradient
$\triangledown_{z}\log p(\theta,z|y)$ can be extremely costly for problems in which the number of latent variables is large; for instance, problems with millions of observations.






\section{Hybrid Unadjusted Langevin}\label{sec:hula}

In this paper we propose a novel ULA method that directly approximates the posterior $p(\theta|y)$. Our method is based on the Langevin diffusion process
\begin{equation}
d\theta_{t} = \triangledown_{\theta}\log p(\theta_t|y)  dt+ \sqrt{2} dB_{t},\label{eq:langevin0theta}
\end{equation}
where $B_t$ is an $m_\theta-$dimensional Brownian motion process. 
The discrete approximation of the ULA corresponding to \eqref{eq:langevin0theta} can be written as
\begin{align}\label{eq:margULA}
\theta_{k+1} & =\theta_{k}+\tau\triangledown_{\theta}\log p(\theta_{k}|y)+\sqrt{2\tau}\epsilon_{k}, \quad \epsilon_{k} \sim N(0_{m_\theta},I_{m_\theta}).
\end{align}
Since the approximation bias of the ULA increases in its dimension, the implied stationary posterior of \eqref{eq:margULA} is subject to less bias than that of \eqref{Eq:Discrete}. Moreover, \eqref{eq:margULA} includes a low-dimensional gradient compared to \eqref{Eq:Discrete}.
To implement the ULA in \eqref{eq:margULA}, one must compute the gradient 
\begin{align}\label{eq:in_int}
\triangledown_{\theta}\log p(\theta|y) = \triangledown_{\theta}\log \int p(y,z|\theta)dz+\triangledown_{\theta}\log p(\theta),
\end{align}
which requires evaluation of the intractable integral $\int p(y,z|\theta)dz$. We assume that the gradient of the prior $\triangledown_{\theta}\log p(\theta)$ is available in closed form.

We make ULA applicable to econometric models with latent variables by constructing an unbiased estimate of the gradient $\triangledown_{\theta}\log p(\theta|y)$, without evaluating the intractable integral in \eqref{eq:in_int}. To this end, we employ the Fisher identity, as used in for example \citet{poyiadjis2011particle}, which shows that the gradient of the marginal likelihood can be expressed as
\begin{align}\label{Eq:Fisher}
   G_{\theta}= \triangledown_{\theta}\log \int p(y,z|\theta)dz =  E_{p(z|y,\theta)} \left[\triangledown_{\theta}\log p(y,z|\theta)\right] .
\end{align}
While the Fisher identity also involves an intractable integral, it allows us to construct an unbiased estimate of it as
\begin{align}
    \widehat{G}_{\theta} =& \frac{1}{S}\sum_{s=1}^S\triangledown_{\theta}\log p(y,z^{(s)}|\theta),
\end{align}
where $z^{(s)}\sim p(z|y,\theta)$ is a draw from the exact conditional posterior distribution of the latent variables. This draw can be produced either using exact Monte Carlo simulation or using an exact MCMC sampling scheme. For most econometric models, these draws can be readily produced. For instance, the states in state space models can be sampled via a Metropolis-Hastings step. The latent utilities in the multinomial probit model, which is a limited dependent variable model, can be generated one at a time via a Gibbs sampler. Regression models with random coefficients allow for exact sampling of the random coefficients.

The proposed sampling method makes ULA for $p(\theta|y)$ feasible by incorporating exact sampling from $p(z|\theta,y)$ to evaluate an unbiased estimate of the gradient $G_{\theta}$.
We refer to our method as the Hybrid Unadjusted Langevin Algorithm (HULA). The step size $\tau$ in HULA trades-off speed of convergence for accuracy to $p(\theta|y)$. 
In contrast to ULA applied to $p(\theta,z|y)$, HULA does not incur any  approximation error related to the posterior of the latent variables. 
Hence, HULA has the potential of faster convergence than exact MCMC methods, with a small loss in accuracy.

\subsection{Subsampling}
HULA generally requires a relatively small number of iterations, making it faster than exact MCMC. However, both methods have to generate from $p(z|y,\theta)$, which can be computationally costly when the number of latents in $z$ is large. For certain types of models, the computational burden of HULA can be further reduced via the use of subsampling techniques.

For models where one can write $p(z|\theta) = \prod_{i=1}^n p(z_i|\theta)$, we have that
\begin{align}
    \triangledown_{\theta}\log p(y,z|\theta)=\sum_{i=1}^n \triangledown_{\theta}\log [p(y_i|z_i,\theta)p(z_i|\theta)],
\end{align}
which has a computational cost that increases linearly in $n$. Our method allows us to reduce the computational burden by considering the representation
\begin{align}\label{Eq:SubsamplingExp}
    \triangledown_{\theta}\log p(y,z|\theta)= E_{f(A)}\left[\frac{n}{M} \triangledown_{\theta} \log p(y_A,z_A|\theta)\right],
\end{align}
where $A\sim f(A)$ denotes a random subset of indexes from $\{1,\dots,n\}$, sampled without replacement. By introducing \eqref{Eq:SubsamplingExp} into \eqref{Eq:Fisher}, we can show that a subsample unbiased estimate of $G_\theta$ can be written as
\begin{align}
    \widehat{G}_{\theta}^M =& \frac{1}{S}\sum_{s=1}^S\frac{n}{M} \triangledown_{\theta} \log p(y_{A^{(s)}},z_{A^{(s)}}^{(s)}|\theta),
\end{align}
where ${A^{(s)}}\sim f(A)$ and $z_{A^{(s)}}^{(s)}\sim p(z_{A^{(s)}}|y_{A^{(s)}},\theta)$.

\subsection{Choice of $\tau$ and $S$}\label{sec:choice}
As mentioned earlier, larger values of $\tau$ lead to faster numerical convergence of the Markov chain, while smaller values lead to slow convergence. However, past a certain step size value, the chain becomes non-ergodic, and hence numerically unstable \citep{hodgkinson2021implicit}.Larger values of $\tau$ also increase the bias of the approximation \citep{vollmer2016exploration}. In our empirical application, we find good performance with $\tau = \frac{1}{n}$, which guarantees smaller steps with more concentrated posterior distributions.

Another caveat to choosing the step size are differences in the magnitude of the marginal posterior variances across parameters. A scalar step size must be small enough to produce stable draws for the parameter with the smallest posterior variance \citep{girolami2011riemann}. This renders the chain to be inefficient for model parameters with large posterior variance, as it becomes slow at traversing the parameter space.
To alleviate this, works such as \cite{roberts2002langevin} and \cite{welling2011bayesian} have suggested the use of a preconditioning matrix. We follow suit in this paper and re-express the ULA step as
\begin{align}
\theta_{k+1} & =\theta_{k}+\tau U\triangledown_{\theta}\log p(\theta_k|y)+\sqrt{2\tau }U^{\frac{1}{2}}\epsilon_{k},\label{Eq:HULAprecond}
\end{align}
where $U$ is a diagonal matrix, in which larger diagonal elements correspond to parameters expected to have a larger posterior variance. 

Our empirical application demonstrates that by setting the number of draws to evaluate the gradient in HULA as $S=1$, we can achieve an accurate approximation while keeping the computational costs low. Algorithm~\ref{alg:hlsa} outlines the HULA steps for generating from the approximation, which we denote as $q_{\tau}(\theta)$.

\begin{algorithm}
    \begin{algorithmic}[1]
        \State{Set a value for  $\tau$, $U$  and initialize $\theta_1$}
        \For{$k=1,\dots,K$}
        \State{Draw $z^{(k)} \sim p(z|y,\theta_k)$}
        \State{Construct $\hat{G}_{\theta_k}=\triangledown_{\theta}\log p(y,z^{(k)}|\theta_k)$}
        \State{Compute $\widehat{\triangledown_{\theta}\log p(\theta_k|y)} = \hat{G}_{\theta_k}+\triangledown_{\theta}\log p(\theta_k)$}
        \State{Draw $\epsilon_{k}\sim N(0_{m_\theta},I_{m_\theta})$}
        \State{Update $\theta_{k+1}  =\theta_{k}+\tau U\widehat{\triangledown_{\theta}\log p(\theta_k|y)}+\sqrt{2\tau}U^{\frac{1}{2}}\epsilon_{k}$}
        \EndFor
    \end{algorithmic}
    \caption{Hybrid Unadjusted Langevin Algorithm}
    \label{alg:hlsa}
\end{algorithm}

\section{Example: Multinomial probit model}\label{sec:mnp}
The multinomial probit (MNP) model is an example of an econometric model for which MCMC sampling is time-consuming. Evaluation of the likelihood function involves analytical integration of a large number of latent variables, which is infeasible. Thus, MCMC samples from the augmented posterior distribution, which requires generation of all latent variables at each iteration and induces high autocorrelation in the Markov chain. In addition, the MNP specification we consider here requires the use of blocked random walk Metropolis-Hastings algorithms for sampling of the model parameters, which exacerbates the problem of high autocorrelation in the chain.

HULA speeds up estimation of the MNP model in three ways. First, because it uses the gradient of the logarithm of the augmented posterior density, its chain quickly converges towards regions of high posterior probability.
Second, independent of the specification of the model, in HULA the parameters are generated all at once rather than in multiple blocks. Third, HULA allows for subsampling of the latents in each sample iteration, which leads to a substantial increase in computational efficiency. 

\subsection{Model specification}
We observe a multinomial choice $y_i$ for individual $i=1,\dots,n$, where $y_{i}=j$ if individual $i$ chooses choice alternative $j=0,1,\dots,J$. Let ${z}_{i}=(z_{i1},\dots,z_{iJ})^\top$ be a $J$-dimensional vector of continuous random variables representing the latent utilities for the choice alternatives, which excludes the base-category latent utility $z_{i0}=0$.

The multinomial outcome $y_{i}$ is determined by the maximum value of ${z}_{i}$:
\begin{align}\label{eq:Y_i}
    y_{i} = \begin{cases} 0 & \text{ if } \max({z}_{i})<0,\\
    j & \text{ if } z_{ij}=\max({z}_{i})>0,\end{cases}
\end{align}
where $\max({z}_{i})$ is the largest element of ${z}_{i}$. The latent utilities corresponding to the choice alternatives are modeled as
\begin{align}\label{eq:Z_i}
    {z}_{i} = X_{i}{\beta} +{\varepsilon}_{i}, \quad {\varepsilon}_{i}\sim N({0}_J,\Sigma),
\end{align}
where $X_{i}$ is a $J \times r$ regressor matrix, ${\beta}$ is an $r$-dimensional vector of coefficients, and ${\varepsilon}_{i}=(\varepsilon_{i1},\dots,\varepsilon_{iJ})^\top$ is a $J$-dimensional normally distributed disturbance vector with mean zero and covariance matrix $\Sigma$.

We consider a factor structure for $\Sigma$. Define the $J\times p$ matrix $B$ with $p \leq J$ and the $J\times J$ diagonal matrix $D$. We model $\Sigma$ as 
\begin{align}\label{eq:factor}
    \Sigma = BB^\top+D^2.
\end{align}
The total number of parameters in $B$ and $D$ is $J(p+1)$. This implies that for a given value of $p$, the number of parameters grows linearly with $J$, instead of quadratically.

Since the scale of the covariance matrix of the latent utilities is not identified in a multinomial probit model, we follow the approach of \citet{loaiza2021scalable} and transform the elements of $B$ and $D$ into a spherical coordinate system. This system is parametrized by a $(J(p+1)-1)$-dimensional vector of angles ${\kappa}$, where the covariance matrix $\Sigma=\Sigma({\kappa})$ is constructed from ${\kappa}$.

The augmented likelihood function of the model is given by
\begin{align}
  p(y,z|X,{\theta}) & = p(y|z) p(z|X,{\theta})= \prod_{i=1}^{n}p(y_i|{z}_i) \phi_{J}\left({z}_i;X_i{\beta},\Sigma({\kappa})\right),
\end{align}
where $\theta = (\beta^\top,\kappa^\top)^\top$, $\phi_{J}\left({z}_i;X_i{\beta},\Sigma({\kappa})\right)$ denotes the $J$-variate normal density with mean $X_i{\beta}$ and covariance matrix $\Sigma({\kappa})$, $y = (y_1,\dots,y_n)^\top$, $z = (z_1^\top,\dots,z_n^\top)^\top$, $X=(X_1^\top,\dots,X_n^\top)^\top$, and 
\begin{align}\label{eq:y_ik}
    p(y_{i}|{z}_{i}) = \begin{cases} I\left[z_{iy_{i}}=\text{max}({z}_{i})\right]& \text{ if } \text{max}({z}_{i})>0,\\
    I(y_{i}=0) & \text{ if } \text{max}({z}_{i})\le0 ,\end{cases}
\end{align}
where $I[A]$ is an indicator function that equals one if $A$ is true and zero otherwise.

\subsection{Estimation}
We are interested in estimating the posterior density 
\begin{equation}
     p({\theta}|{y},X)\propto p({\theta})\int p({y},z|X,{\theta}) dz, \label{Eq:exact_posterior}
\end{equation}
with prior $p({\theta}) = p(\beta)p(\kappa)$,  $p({\beta}) = \phi_{r}({\beta};0_r,\frac{1}{10}I_r)$, and $p(\kappa)$ has an implied prior mean for $\Sigma$ that equals the equicorrelated covariance matrix $\Sigma_{\text{equi}}=\frac{1}{2}(I_J + \iota_J \iota_J^\top)$, as specified in \citet{loaiza2021scalable}.

We use HULA to construct an approximation to this posterior, which requires an analytical expression for $\triangledown_{\theta}\log p({y},{z}|X,{\theta}) p({\theta})$, a draw from the full conditional distribution of the latent utilities $p({z}|{\theta},{y},X)$, and a choice for the preconditioning matrix $U$ and the step size $\tau$. 
First, \citet{loaiza2022fast} derives an expression for the gradient. 
Second, the latent utilities are sampled according to the Gibbs sampling algorithm proposed by \citet{geweke1991efficient}. Third, we set $\tau=\frac{1}{n}$, the diagonal elements of $U$ corresponding to ${\beta}$ equal 0.99 divided by the diagonal elements of the matrix $\frac{1}{n}\sum_{i=1}^n X_i^\top \Sigma_{\text{equi}} X_i$, and the diagonal elements of $U$ corresponding to ${\kappa}$ equal $0.1$.

The HULA predictive probability mass function for $y_i$ is given by
\begin{align}\label{eq:predictive}
p_{\tau}({y}_i|X_i,y,X) = \int p(y_i|z_i)\phi_{J}\left({z}_i;X_i{\beta},\Sigma({\kappa})\right)q_{\tau}(\theta)dz_id\theta,
\end{align}
where $X_i$ denotes the attributes of the observation $i$ to be predicted. An estimate $\hat{p}_{\tau}({y}_i|X_i,y,X)$ for the predictive in \eqref{eq:predictive} is constructed as the empirical probability mass implied by the draws ${y}_i^{[k]}$ based on ${z}_i^{[k]}$, with ${z}_i^{[k]}\sim \phi_{J}\left({z}_i;X_i\beta^{[k]},\Sigma(\kappa^{[k]})\right)$ where $\{\beta^{[k]}\}_{k=1}^K$ and $\{\kappa^{[k]}\}_{k=1}^K$ denote the parameter draws from $q_{\tau}(\theta)$.

\subsection{Pasta purchases}
We fit the multinomial probit model to a data set on pasta brand purchases. This consumer choice data set includes one million purchases and is made available by the Dunnhumby data platform\footnote{https://www.dunnhumby.com/source-files/} as ``Carbo-Loading: A Relational Database".  The final sample includes purchases without coupons of the ten top-selling pasta brands, excluding private labels, and contains 1,070,436 observations. We randomly allocate 80\% of the observations for estimation of the model, and the remaining 20\% are employed for out-of-sample evaluation. 
The data set is described in detail by \citet{loaiza2022fast}, who fit the multinomial probit model discussed above with an intercept and the log price for each brand with VB. We follow the same model specification and compare our method to exact MCMC and VB.


\subsubsection{Computational efficiency}
This section compares the computational efficiency of the proposed HULA method to that of the exact MCMC sampler for the multinomial probit model.

First, per sample iteration, the computational cost of HULA is lower than that of MCMC: 1.64 versus 1.81 seconds, respectively\footnote{The methods are implemented in a HP Z240 SFF Workstation with an Intel i7-7700 CPU at 3.6GHz}. Both samplers generate at each iteration the latent utilities for all pasta brands and all observations from its full conditional distribution. MCMC samples the coefficients with a Gibbs step and the angles with a blocked random walk Metropolis-Hastings step. Instead, HULA produces a new of draw of all the parameters at once by evaluating the gradient of the augmented posterior, which is less time-consuming.

\begin{figure}[tb!]
\caption{Trace plot across iterations of the HULA and exact MCMC chain.}
\centering
\includegraphics*[width=\textwidth,trim = 0 0 0 0]{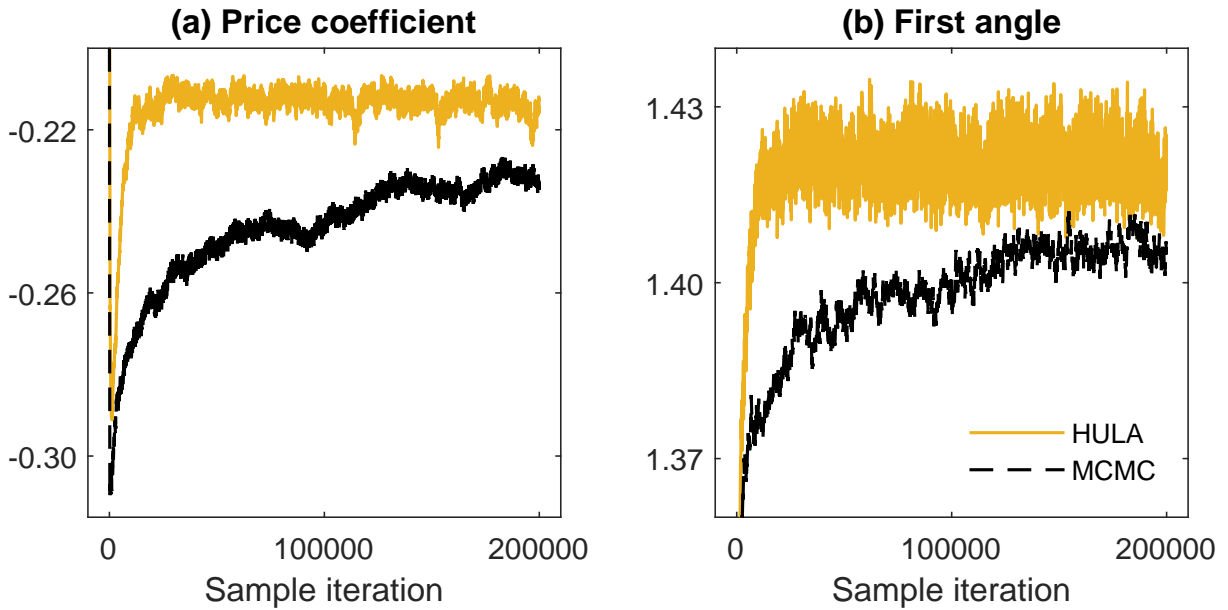}\\[3mm]
\fnote{This figure shows the trace plot of the price coefficient $\beta_{10}$ (Panel (a)) and the first angle $\kappa_1$ (Panel (b)) for the first 200,000 iterations of HULA (solid yellow line) and exact MCMC (dashed black line).}
\label{fig:pasta_conv}
\end{figure}

Second, the HULA chain converges substantially faster than that of MCMC. Figure~\ref{fig:pasta_conv} shows the trace plot of the price coefficient and the first angle parameter of the first 200,000 iterations. For both parameters, the HULA chain converges well within the first 100,000 iterations. However, the MCMC chains have still not converged to the region of high posterior probability at 200,000 iterations. We find similar patterns in the chains for the other parameters: the 9 intercepts in $\beta$ and the 16 other angles in $\kappa$. Based on the trace plots, we conclude that the HULA chain has converged after 100,000 iterations, and the MCMC chain after 1,000,000 iterations.

After discarding these burn-in iterations, we use the remaining iterations to construct our results. For HULA we collect 100,000 draws after the burn-in, and for MCMC 1,000,000 draws. 

\begin{figure}[tb!]
\caption{Effective sample size per iteration of HULA relative to exact MCMC}
\centering
\includegraphics*[width=\textwidth,trim = 0 0 0 0]{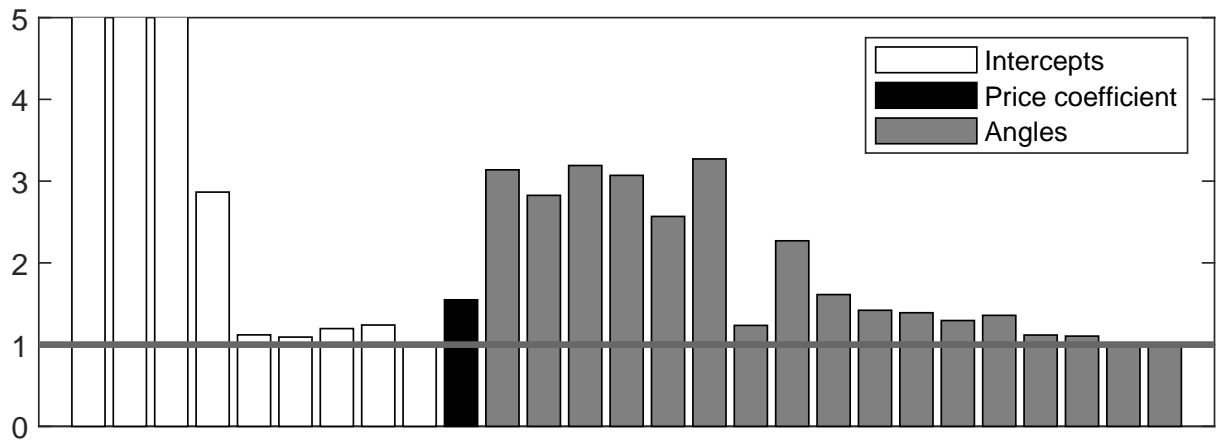}\\[3mm]
\fnote{This figure shows the effective sample size per iteration of HULA relative to exact MCMC, for the intercepts (white bars), price coefficient (black bar), and angles (gray bars). Values larger than 1 favour HULA over exact MCMC. The effective sample size is calculated with the autocorrelation lag order truncated at 1000 lags. The relative effective sample size per iteration for the first three coefficients are truncated at 5.}
\label{fig:pasta_ess}
\end{figure}

We assess the quality of the collected samples by the effective sample size calculated with an autocorrelation order of 1000 lags. Figure~\ref{fig:pasta_ess} shows the effective sample size per iteration of HULA relative to MCMC. The white, black and grey bars correspond to the intercept coefficients, price coefficient, and angle parameters, respectively. Since values larger than 1 favour HULA over MCMC, HULA obtains a higher effective sample size per iteration for all parameters. This result, in conjunction with the fact that HULA is also faster per iteration, and its chain reaches convergence in fewer iterations than MCMC, make it much more computationally efficient in this example.

The HULA approach takes in total 91 hours. This is still more computationally costly than the 15 hours of computation time required by VB. However, as we will show below, HULA is more accurate and the difference in computational costs can be reduced substantially by constructing the gradient using subsampling.

\subsubsection{Posterior accuracy}
In the MNP model, the parameter estimates themselves are hard to interpret and as such are not considered to be the key output from the model. Instead, in most empirical applications practitioners are interested in the implied choice probabilities of the alternatives.
Thus, to assess the accuracy of the proposed HULA method, we compare their fitted posterior choice probabilities to those estimated by the exact MCMC method and the approximate VB method. 

\begin{figure}[tb!]
\caption{Purchase probabilities for two pasta brands}
\centering
\includegraphics*[width=\textwidth,trim = 0 0 0 0]{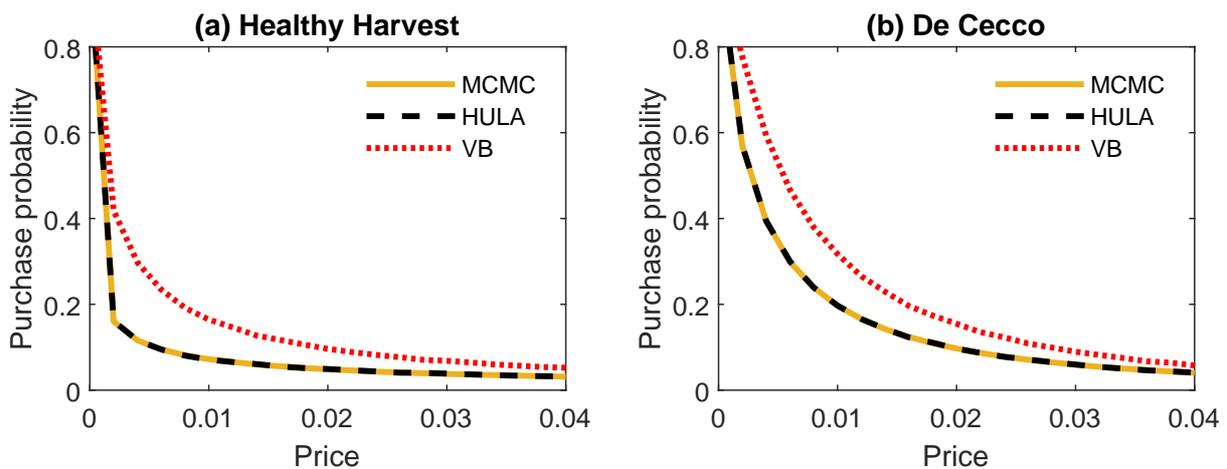}\\[3mm]
\fnote{This figure shows the posterior purchase probabilities of the pasta brand `Healthy Harvest' (Panel (a)) and `De Cecco' (Panel (b)) as a function of their price, with the prices of the other brands fixed at their mean. The solid yellow line shows the probabilities estimated with exact MCMC, the dashed black line with HULA, and the dotted red line with VB. }
\label{fig:pasta_prob}
\end{figure}

We find that HULA accurately estimates the posterior choice probabilities. Figure~\ref{fig:pasta_prob} shows the choice probability of buying the pasta brand `Healthy Harvest' (Panel (a)) and `De Cecco' (Panel (b)) as a function of their price, with the prices of the other pasta brands fixed at their mean. The solid yellow lines correspond to the posterior probabilities of MCMC, the dashed black lines to HULA, and the dotted red lines to VB. 
HULA produces posterior probabilities that are almost identical to MCMC. We find the same for the purchase probabilities of the other 8 brands. However, Figure~\ref{fig:pasta_prob} shows that for `Healthy Harvest' and `De Cecco' VB is less accurate. We find that for the most popular pasta brands, the three methods have similar posterior purchase probabilities.

\subsubsection{Predictive accuracy}
HULA also attains similar predictive accuracy to MCMC. Table~\ref{tab:forecast_pasta} shows  the in- and out-of-sample log-scores and hit-rates of HULA, VB, and exact MCMC. Larger values for these metrics are preferred. The log-scores and hit-rates of HULA and MCMC are similar. However, although the differences are small, HULA performs better on all four metrics relative to Variational Bayes. 

\begin{table}[tb!]
  \centering
  \caption{Predictive assessment}
  \begin{threeparttable}
    \begin{tabular}{llrrrrr}
    \toprule \toprule
          &       & \multicolumn{1}{l}{VB} & \multicolumn{1}{l}{HULA(20\%)} & \multicolumn{1}{l}{HULA} & \multicolumn{1}{l}{MCMC} & \multicolumn{1}{l}{Naive} \\ \midrule
    \multicolumn{1}{l}{in} & log-score & -1.77322 & -1.77305 & -1.77298 & -1.77306 & -1.84478 \\
    \multicolumn{1}{l}{in} & hit-rate & 0.39802 & 0.39869 & 0.39888 & 0.39882 & 0.29503 \\
    \multicolumn{1}{l}{out} & log-score & -1.77410 & -1.77379 & -1.77377 & -1.77375 & -1.84235 \\
    \multicolumn{1}{l}{out} & hit-rate & 0.39656 & 0.39674 & 0.39678 & 0.39689 & 0.29601 \\ 
          \bottomrule \bottomrule
    \end{tabular}%
\begin{tablenotes}
\footnotesize
\item This table shows the in- and out-of-sample log-scores and hit-rates, defined in, for instance, \cite{loaiza2022afast}. The final row shows the total estimation time in hours. Predictive densities are estimated with VB, HULA with 20\% subsampling, HULA with no subsampling, exact MCMC, and a naive method in which the forecast equals the most frequently observed category.
\end{tablenotes}
\end{threeparttable}
  \label{tab:forecast_pasta}
\end{table}

\subsubsection{Hybrid Unadjusted Langevin with subsampling}
The computational efficiency of HULA can be improved by constructing an unbiased estimate for the gradient at each iteration with a random subsample of observations, which implies that only a subsample of latent utilities must be generated.
We run HULA with 20\% subsampling, which takes 22 hours. This is a substantial reduction in computation time relative to HULA without subsampling (91 hours). Moreover, the computational cost is of the same order as that of VB (15 hours).

The loss in posterior accuracy due to subsampling is small. We find that the posterior probabilities of HULA with 20\% subsampling are visually indistinguishable from those of HULA without subsampling, and are thus not included in Figure~\ref{fig:pasta_prob}. Hence, both HULA with and without subsampling produce posterior purchase probabilities that are very close to the exact posterior purchase probabilities of MCMC.
The predictive accuracy of the HULA with subsampling is also similar to HULA without subsampling. Table~\ref{tab:forecast_pasta} shows that the in- and out-of-sample log-scores and hit-rates of HULA with 20\% subsampling and MCMC are still alike, and that HULA with 20\% subsampling still outperforms VB.

We conclude that the reductions in computation time due to subsampling are substantial, while the loss in accuracy is small.

\section{Conclusion}\label{sec:conclusion}
This paper proposes an approximate Bayesian estimation method that is based on the unadjusted Langevin algorithm. This method is gaining popularity in the machine learning literature due to its simplicity and its convenient convergence properties. We show how the method can be employed to produce accurate estimation of latent variable models with large data sets, which is a challenging estimation problem in econometrics. 

We refer to our method as the hybrid unadjusted Langevin algorithm (HULA). The approach targets the marginal posterior distribution of the model parameters, where an unbiased gradient estimate for the logarithm of the marginal posterior density is constructed via the use of the Fisher identity. 
The advantages of HULA are illustrated in a discrete choice application with big data: HULA produces fast convergence, negligible loss in posterior and predictive accuracy, and a substantial reduction in computational costs. 


\bibliographystyle{apalike} 
\bibliography{langevin}

\end{document}